\documentclass[aps,prl, superscriptaddress,showpacs,floats,twocolumn,amssymb,floatfix]{revtex4-1}

\usepackage{bm}
\usepackage{mhchem}
\usepackage{amsmath,stmaryrd,MnSymbol,amssymb}
\usepackage{graphicx}
\usepackage[usenames,dvipsnames]{color}
\usepackage[colorlinks,citecolor=magenta]{hyperref}

\begin{document}
\addtolength{\topmargin}{11pt}

\title{Atlas of 2D metals epitaxial to SiC: filling-controlled gapping conditions and alloying rules}
\author{Yuanxi Wang}
\email{yow5110@psu.edu}
\affiliation{Materials Research Institute, Pennsylvania State University, University Park, Pennsylvania 16802}
\affiliation{2-Dimensional Crystal Consortium,Pennsylvania State University, University Park, Pennsylvania 16802}

\author{Vincent H. Crespi} 
\email{vhc2@psu.edu}

\affiliation{Department of Physics, Pennsylvania State University, University Park, Pennsylvania 16802, USA}
\affiliation{Department of Chemistry, Pennsylvania State University, University Park, Pennsylvania 16802, USA}
\affiliation{Department of Materials Science and Engineering, Pennsylvania State University, University Park, Pennsylvania 16802, USA}
\affiliation{2-Dimensional Crystal Consortium,Pennsylvania State University, University Park, Pennsylvania 16802}
\affiliation{Material Research Institute, Pennsylvania State University, University Park, Pennsylvania 16802}

\date{\today}

\begin{abstract}
The realization of air-stable 2D metals epitaxial to SiC and capped by graphene creates a potentially immense chemical space of 2D metals and alloys that could expand the variety of solid-state excitations unique to 2D metals beyond what is known for graphene and niobium/tantalum chalcogenides. We perform a high-throughput computational survey from first-principles predicting the structures and stabilities of all metals in the periodic table when they intercalate graphene/SiC. Our results not only agrees with all experimentally known metal/SiC structures explored so far, but also reveals conspicuous trends related to metal cohesive energies and metal-silicon bonding. For special groups of metals, a small bandgap opens, relying on appropriate electron filling and substrate-induced symmetry breaking. From this gapping stabilization, we derive alloying rules unique to 2D metals.\end{abstract}
\maketitle 

Metals make up $\sim$80\% of the periodic table, yet their representation in the ever-expanding family of atomically thin 2D materials is very limited due to their susceptibility to surface oxidation \cite{Kochat2018, Xing2015}. This fragility applies even to metallic compounds such as \ce{NbS2} and \ce{NbSe2}, which require encapsulation to prevent device degradation \cite{Yan2019, Xi2016}. Limited in their compositional variety as they may be, 2D metals such as niobium dichalcogenides already support a rich variety of solid-state excitations unique to 2D, such as exotic pairing  in 2D superconductivity \cite{Xi2016} and near-zero dispersions in 2D plasmons \cite{DaJornada2020}. The recent realization of air-stable elemental 2D metals has established a scalable synthesis protocol generalizable to a wide range of elements, where metals intercalate into a high-energy interface -- graphene/SiC -- and are subsequently sealed under healed graphene. 
Successful intercalants so far include Ga, In, Sn \cite{Briggs2020}, Ag \cite{Rosenzweig2020}, Cu \cite{Forti2016}, Ca  \cite{Kotsakidis2020}, Mg \cite{Grubisic-Cabo2020}, and Au \cite{Forti2020}. Many other metals intercalating graphene/SiC (Pt \cite{Xia2014}, Pb \cite{Yurtsever2016}, Mn \cite{Gao2012}) or depositing on SiC (Bi \cite{Reis2017}) only under ultra-high vacuum conditions also awaits the trial of scalable intercalation. 
Since metal intercalation is mainly driven by replacing a lattice-mismatched SiC-graphene interface with a lattice-matched (and energetically more favorable) SiC-metal interface, intercalation can be anticipated for even more metallic elements. The chemical space of possible 2D metals thus easily surpasses the 2D metals known so far, and has already led to interesting physical phenomena such as coexistence of two spin textures \cite{Yaji2019}, Dirac fermion and quantum spin Hall systems \cite{Li2013a}, extended van Hove singularities \cite{McChesney2010, Link2019}, 2D heavy metals for valleytronics \cite{Xu2017}, 2D superconductors \cite{Briggs2020, Zhang2010, Xing2015, Zhang2015a}, and large optical nonlinearities \cite{Steves2020}. 2D metals protected at the graphene/SiC interface also hold promise as a platform for topological superconductivity, since they can provide pristine large-area surfaces to interface with a topological insulator. Considering 2D alloys opens up an even larger chemical space, since (entropy-driven) alloy formation is typically not constrained by e.g. an octet rule. Alloying would allow fine-tuning of physical parameters required to achieve exotic phases or optimal device performance, e.g. tuning the average electron count per atom in search of a maximized $T_c$ in superconducting alloys (Matthias' rules \cite{Webb2015}). This intriguing chemical space prompts an investigation of the fundamental properties of 2D metals, including metallicity, stability, and alloy rules. 
Relating metallicity to dimensionality is especially intriguing. For 3D metals, it is well-known that, due to strong band overlap near the Fermi level, the metallicity of an element is not determined by (a naive picture of) electronic bands being filled or half-filled with even or odd number of electrons; metallicity is often instead determined by the so-called Herzfeld criterion \cite{Herzfeld1927}. 
\begin{figure*}
\centering
\includegraphics[width=0.9\textwidth]{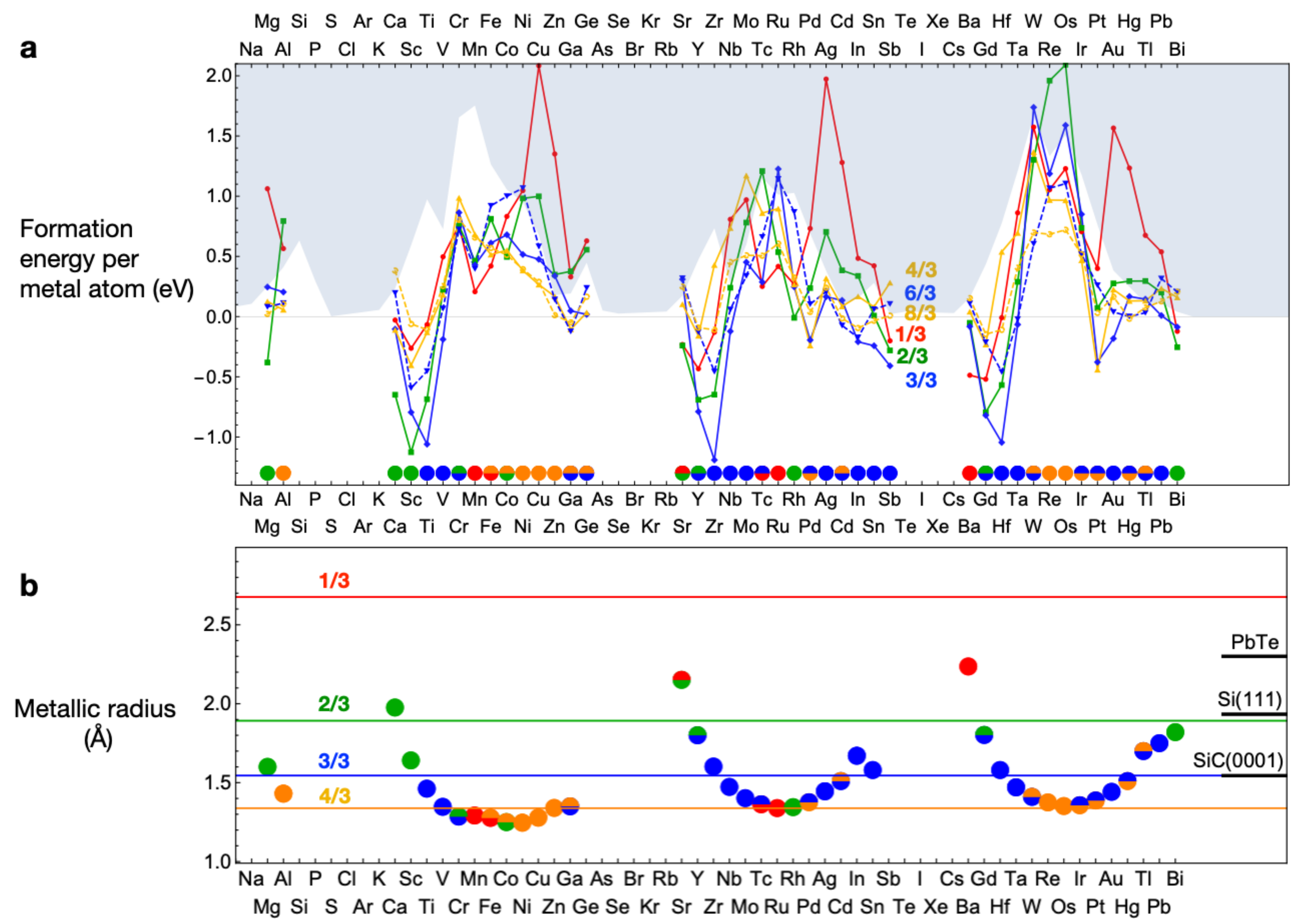}
\caption{(a) Formation energy per atom for all 2D metals, referenced to bulk 3D ground states. For comparison, estimated per-atom energies for metal surfaces are given by the filled curved in the backdrop. Colored circlular markers on the bottom indicate the predicted ground state of 2D metals. A second color indicates a metastable phase within +0.1~eV. (b) Same markers as in (a) but now attached to metallic radii data for each element examined. Horizontal lines indicate average interatomic distances for each of the four monolayer metal structures: 1/3, 2/3, 3/3, and 4/3 in red, green, blue, and orange.}
\label{all} 
\end{figure*}
At low dimensions, developing a bandgap or pseudo gap within band theory (i.e. Wilson transition) becomes possible with the help of lower coordination \cite{Dowben2000} and substrate interaction. Even if we do not require the developed gap to be at the Fermi level, local gaps in the Brillouin zone of 3D metals (which are not uncommon, e.g. at Brillouin zone boundaries of alkali metals) are less likely to overlap and yield a global gap across the entire Brillouin zone than in 2D metals.
Known precedents of gapping 2D metals include monoelemetals alkali earths \cite{Yakovkin1999, Dowben2000}, group 10 transition metals \cite{Burdett1994}, mercury (see Ref.~\cite{Nevalaita2018}), Au and Ag on SiC \cite{Forti2020, Rosenzweig2020}, and Ag on Si(111) \cite{Ding1991}. Since each study was limited to one element or one group, often stabilized by one particular matching substrate or even in hypothesized free-standing form, no general trends in metallicity have yet been extracted. In this Letter, based on density functional theory (DFT), a systematic high-throughput computational survey of all metal elements that can realistically intercalate graphene/SiC not only reproduces all existing experimentally confirmed structures, but also reveals clear design rules for which systems are gapped on SiC. We show that gapping conditions originate from substrate-induced symmetry breaking and electron filling conditions, and that gap formation affects metal stability and alloying behavior. Conclusions are generalizable to other common substrates such as Si(111) and the (0001) surfaces of zincblende or wurtzite semiconductors. The establishment of generalizable gapping regimes within a wide composition space can thus provide a robust basis for engineering small-gap semiconductors for applications such as infrared detectors or topological insulators, where material parameters (e.g. spin-orbit coupling or electron-phonon coupling) could be tuned with greater freedom than in existing families of 2D materials and alloys.

We first determine ground state structures. For each of the 43 elements considered in Fig.~\ref{all}a, we screen through a pre-defined set of lattice structures based on triangular and hexagonal metal lattices on SiC (0001) and capped under graphene, with different metal thicknesses and lateral densities, based on the expectation that a principle of parsimony and bonding with a hexagonal 6H-SiC substrate would favor these simple lattices for the majority of 2D metals. Structural relaxations are then performed for all trial structures. Since the SiC (0001) surface termination is a triangular lattice of Si, we consider trial metal monolayer structures labeled by their coverage with respect to the Si layer: 1/3, 2/3 (hexagonal), 3/3, and 4/3 (incommensurate) colored in red, green, blue, and orange; for the last two high-density coverages we also consider bilayer trial structures labeled by 6/3 and 8/3 and represented using dashed blue and orange lines. All calculations includes graphene since our main goal is assessing likelihood of intercalation. See the Method section on details of the structural relaxations.

\begin{figure*}
\includegraphics[width=0.9\textwidth]{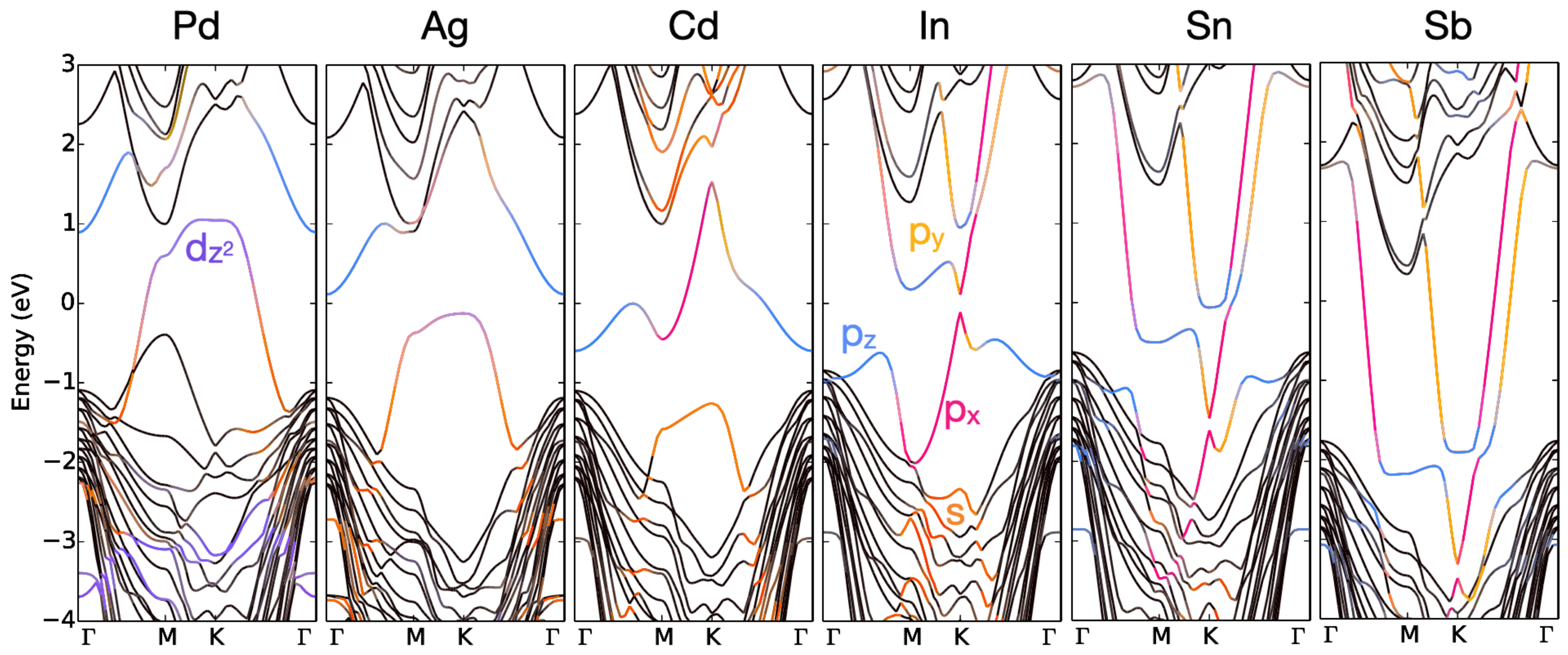}
\caption{DFT band structure of 2D metals on SiC. Wavefunction projection onto metal $s$, $p_{x,y,z}$, and $d_{z^2}$ orbitals are colored in orange, red, yellow, blue, and purple. Notice the similarities between band structures of adjacent elements and the downshift of all metal bands with increasing atomic number.}
\label{bands} 
\end{figure*}
Formation energies plotted in Fig.~\ref{all} are defined by the total energy difference between the intercalated and unintercalated systems, referenced to per-atom bulk metal energies,
\begin{equation}
E_\text{form} = E_\text{intercalated} - E_\text{unintercalated} - E_\text{bulk metal}
\end{equation}
Surveying the stabilities of all metals in Fig.~\ref{all}a, several trends emerge. 
The first \textit{energetic} trend concerns the overall arch shapes of the formation energy curves in each period, resembling those of metal cohesive energies: in each period, it is well-known that metal cohesive energies \cite{burdett1995chemical} (and metal surface energies) increase from early- to mid-period elements (Mn, Tc, Re) as bonding orbitals are filled and decrease towards late-period elements as anti-bonding orbitals are filled (see also Friedel's model \cite{ziman2011physics}). The calculated metal/SiC formation energies mostly follow along while staying below metal surface energies (taken from Ref.~\cite{Tran2016a}, shown in backdrop). This downshift represents the energetic preference of metal-silicon bonding over metal-metal bonding, since we can consider metal/SiC structures and metal surfaces as 2D metal on SiC and 2D metal on metal respectively. The downshift is not a rigid one: structures involving elements near Ti/Zr/Hf and near Co/Rh/Pt stabilize the most, while ones involving elements near Cr/MoW  stabilize the least. This varying degree of stabilization closly follows the variations in the formation energies of metals forming silicides (see Supplementary Information), a descriptor that captures the energetic preference of metal-silicon bonding over metal-metal bonding. The formation energies of \textit{some} structures being positive, referenced against bulk 3D parents, does not necessarily suppress prospects of intercalation, since, before reaching thermodynamic endpoints of bulk metals, metal vapor would have to at least go through 2D or nanoparticle phases deposited on graphene to prepare for intercalation, thereby incurring a high nucleation energy. A lower-bound estimate of this barrier of nucleation on graphene can be given, again,  by metal surface energies in the background of Fig.~\ref{all}a. Compared with these metal surface energies, all metal/SiC formation energies are shown to be more negative, suggesting that intercalation is likely.

One minor exception to the above trends is the 1/3 coverage preference for Mn and Tc, where the formation energy dips towards zero. In a previous report where the local density approximation was used \cite{Li2013a}, Mn intercalation was even predicted to be favorable.  This low-coverage preference is due to achieving a special half-metal electronic structure at exactly the band filling of Mn \cite{Li2013a} (and presumably Tc). Another interesting exception is the dips at Pd and Pt; we will return to this in the discussion on electronic structure. Unpredicted complications include possibilities of forming silicides, i.e. not simple metal-on-silicon architectures but surface layers with mixed metal and silicon atoms and structural motifs similar to 3D silicides. 2D silicide structures are more likely to be favored by elements before group 10, as can be inferred from phase diagrams (see SI) and experimentally confirmed cases of \ce{Ni2Si} and \ce{MnSi}. 
2D silicides may require \textgreater 1300K temperatures to convert from SiC due to the stability of SiC \cite{Yaney1990}.

Also conspicuous is a \textit{structural} trend: ground-state 2D metal structures are mainly governed by their native lattice constants. In every period, metallic radii \cite{Ong2013, burdett1995chemical} first decreases with increasing group number from $\sim$2.5~\text{\AA} to $\sim$1.3~\text{\AA} and then increases to $\sim$1.5~\text{\AA}, as shown in Fig~\ref{all}b. The average inter-atomic distances of the four types of monolayer structures shown in the same figure all fall within the range of possible metallic radii. The actual structures adopted by the 2D metals -- color-coded in red for 1/3, green for 2/3, blue for 3/3, and orange for 4/3 -- clearly shows that they are determined mainly by native metallic radii, suggesting a moderate independence against lattice matching SiC. Thus if a substrate with a lateral lattice constant larger than that of SiC (0001) is considered (see right boundary of Fig.~\ref{all}b), the preferred 2D metal density should be preserved, meaning nominally denser structures relative to a now-diluted layer of substrate atoms. This expectation is consistent with Pb favoring a 4/3 phase and In favoring a 6/5 phase when grown on Si(111) \cite{Zhang2010}.
The majority of preferred structures is 3/3. Whether this is due to the SiC(0001) lateral lattice constant happening to coincide with the greatest number of metallic radii or due to the simple 3/3 and 6/3 structures being favored (since they lattice-match SiC) is not clear, but could be determined by future computational screening of metals on Si(111). Overall, predicted lattice structures achieve excellent match with prior reports on successful synthesis; a full account is given in the Supplemental Information. Exceptions to the above structural trends include Mn and Tc adopting a 1/3 coverage, as discussed previously. Another exception is Co and Rh adopting 2/3 structures. The 2D metal formation energy for the 2/3 structure (green) dipping at these elements in group 9 and also at groups 3 and 4 correlate well with metals in these exact groups reaching the most negative metal silicide formation energies (see Supplemental Information). Presumably the 2/3 structure more fully exploits the strong metal-silicon bonding in these groups by better allowing metal atoms to bond with silicon rather than with neighboring metals.

The third trend concerns \textit{stacking}: bilayer 6/3 structures (dashed blue) become more stable than monolayer 3/3 (solid blue) for groups 12 and 13 (e.g. Zn and Ga). These are the only elements in the 3/3 phase with partially filled $p_z$ orbitals that can be further stabilized by interacting with $p_z$ orbital in an additional layer.

All the trends and exceptions discussed above originate from the evolution of 2D metal electronic structures across the periodic table. Here we focus on ground-state (or near ground-state) 3/3 structures from Pd up to Sb since 3/3 is most common. Trends observed from Pt to Bi are similar. For simplicity, we only show band structures for structures without graphene in Fig.~\ref{bands}; including graphene would only introduce small amounts of charge transfer. In every band structure, metal bands interacting with Si dangling bonds reside in a large SiC gap. Bands from surface silicon (see orbital projection onto surface Si $p_z$ in SI) achieve the lowest band energies for Pd (hybridizing with metal $d_{z^2}$) and Sb (hybridizing with metal $p_z$), explaining their stabilities shown by the dips in the blue curve in Fig.~\ref{all}a. 
Turning to metal orbital projections, we observe that, with increasing atomic number, bands with $s$, $p_{x,y,z}$, and $d_{z^2}$ characters (in orange, red, yellow, blue, and purple) downshift in an almost rigid fashion, framed within a sizable SiC gap (dense black bands). 

As metal bands parade through the SiC bandgap, local (in reciprocal space) gaps already appearing within the metal bands set up the conditions for creating a global gap. For two cases, Ag and Ga (also Au and In), filled bands reach a global gap.  We examine the prerequisites enabling each of the above two processes: gap formation and appropriate filling. The first ingredient, local gaps, originate from symmetry breaking by the substrate: breaking mirror symmetry in the $z$ direction opens $p_z-p_x$ and $p_z-p_y$ hybridization gaps along $\Gamma-M$ and $\Gamma-K$ respectively (see the tight-binding model in the SI and Ref.~\cite{Petersen2000}) and breaking in-plane reflection symmetries from six mirror planes ($C_{6v}$) to three ($C_{3v}$)  opens a $p_x-p_y$ hybridization gap at K. These statements are further corroborated by group theory analysis. A hypothetical freestanding 2D triangular metal lattice in $D_{6h}$ symmetry at $\Gamma$ transforms into $C_{2v}$ along $\Gamma-M$ and $\Gamma-K$ as enforced by compatibility relations, yielding band crossings between bands with unlike representations; breaking mirror symmetry in the $z$ direction into $C_{6v}$ would transform into $C_s$ along these paths, yielding anti-crossings between bands now in the same representation.  
Further symmetry lowering from $C_{6v}$ to $C_{3v}$ changes the K point group from $C_{3v}$ to $C_3$, and breaks a doubly degenerate representation of $p_x$ and $p_y$ states into two separate 1D representations, yielding a gap. Indeed, detaching the 2D metal from the SiC substrate by 1~\text{\AA} immediately closes the gap at K and reduces the other two gaps.
A similar analysis applies to the case for Ag/SiC, where the bandgap is due to anti-crossings between $s$, $p_z$, and $d_{z^2}$ orbitals. The effect of spin-orbit coupling on heavy metal systems is discussed in the Supplemental Information.

\begin{figure}[h]
\includegraphics[width=0.51\textwidth]{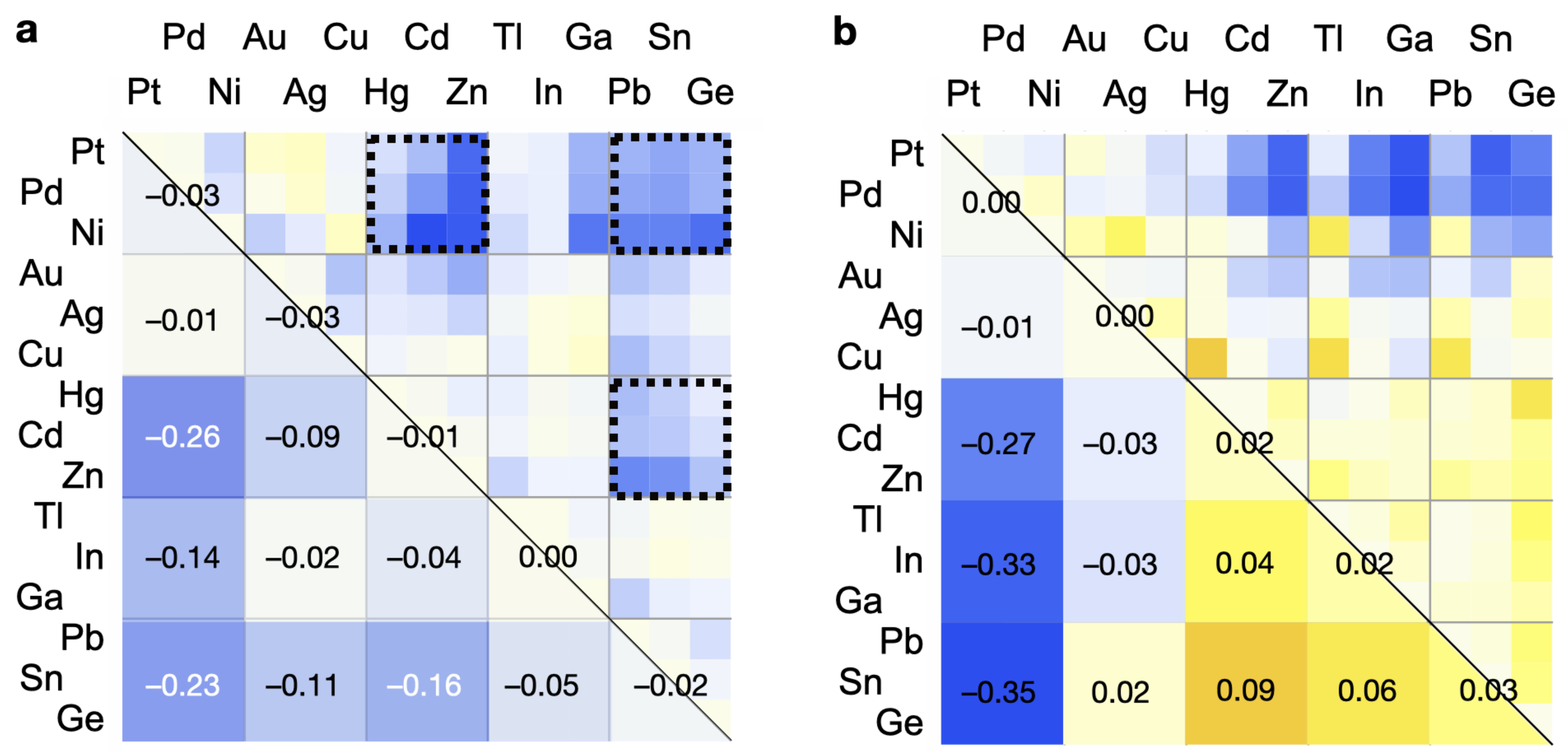}
\caption{Formation energies per metal atom for 1:1 binary alloys (a) in 2D form (on SiC) and (b) in 3D form.  For each chart, the upper-right division tabulates calculated formation energies  and the lower-left division tabulates same-group averages. Clustering patterns in the 2D case is highlighted by dashed lines. }
\label{alloy} 
\end{figure}
The other ingredient for developing a global gap is full occupation of bands below all the overlapping local gaps. Since Si dangling bonds offer one electron each, the condition that one valence band is fully occupied is met when the element contributing to that band also offers one electron. We thus arrive at Ag and Ga meeting this condition with their one $s$ and one $p$ electron. 

\begin{figure*}
\includegraphics[width=0.85\textwidth]{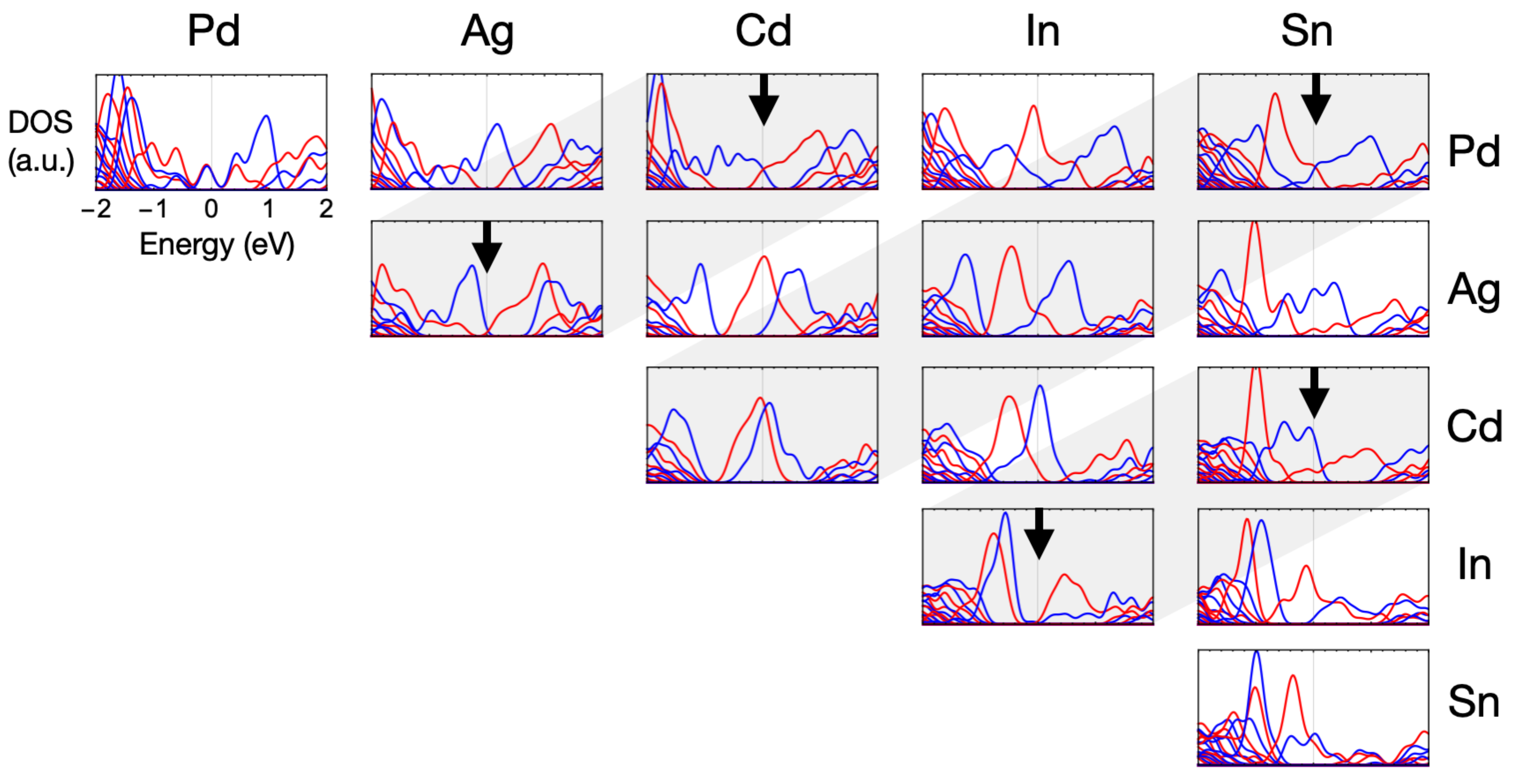}
\caption{Densities of states for a subset of 1:1 binary alloys, in each case decomposed by band index and colored alternatingly with respect to band index in blue and red. Black arrows highlight the cases where the Fermi level coincides with a gap or a dip in the DOS, where alloy formation is preferred. }
\label{dos} 
\end{figure*}
In semiconductor design, the octet rule is a fundamental guiding principle for deriving stable binary semiconductors such as GaN. Similarly, the well-defined gapping conditions for 2D metals suggest that stable 2D alloys can be discovered by isoelectronic substitutions of 2D Ga/In and Ag/Au. We demonstrate such an alloy rule is indeed valid, from a systematic investigation for the 1:1 alloy case. Fig.~\ref{alloy} shows the formation energies per metal atom of all possible binary alloys by pairing among 15 elements from groups 10 to 14. The upper-right division tabulates actual data and the lower-left division tabulates same-group-averaged formation energies. Instead of the expected preference of large electronegativity differences (which would favor the upper-right corner), clustered islands of stability are observed. Isoelectronic to Ag is Pd-Cd and isoelectronic to Ga is Cd-Sn. 
Inspection of their densities of states (DOS, decomposed by band index) in Fig.~\ref{dos} confirms that these two alloys, although not having a gap, are as close as one can be to developing a gap among all alloys considered, with one almost fully filled band and another almost empty band most separated from each other than in other alloys. The third cluster, Pd-Sn, is more complicated: although it is isoelectronic to the half-filled 2D Cd, a rigid band model no longer applies, due to the large electronegativity difference between Pd and Sn. Instead, Pd-Sn stabilize by splitting the half-filled Cd metallic band into one almost filled band and one almost empty band, each with mixed Pd and Sn characters, similar to the case in Ref.~\cite{Pettifor1987}. This separation is apparent when observing the In$\rightarrow$Cd-Ag$\rightarrow$Pd-Sn series.

Comparisons with other known alloy rules are in order. 
The simplest alloy model based on rigid bands and band filling due to Friedel \cite{ziman2011physics, burdett1995chemical} results in alloys simply with larger differences in electronegativities being favored. This is seen for the alloy formation energies for the same set of metals in Fig.~\ref{alloy} (right panel), where formation energies per atom are plotted and they become more negative (more stable) monotonically with increasing electronegativity towards the top-right corner. This monotonicity in 3D alloy stabilities contrasts the clustering pattern seen in 2D alloy stabilities, suggesting a partial binary-compound-like chemical character unique to the 2D alloys.
The filling condition discussed above also share apparent similarities with the Hume-Rothery rule on electron concentration, which states that, when free-electron-like Fermi surfaces intersect Brillouin zone boundaries, deviations from free-electron DOS occur so that fillings up to these DOS features would stabilize the metal (see Refs.~\cite{Paxton1997, Mizutani2016} for a detailed and modern analysis). For 2D metals, the stability requirement is more straightforward: fillings that reach a bandgap are preferred.

The filling condition can be satisfied not only by varying alloy composition, but also by varying the choice of the substrate. For SiC, Si dangling bonds contribute one electron each. Following the rule that each $m$-valent surface atom requires $2-m/4$ electrons to passivate, ZnS (0001) with S termination would require 0.5 electrons from the 2D metal. Indeed, pairing this ZnS surface with a 1:1 Ga-Cd alloy yields a gap (see SI). This shows that gapping 2D metals are not limited to ones grown on SiC, although currently SiC may be the the only substrate that can be prepared into a high-energy interface that allows scalable intercalation.

Overall, our systematic search of 2D metals synthesizable by scalable intercalation and extraction of gapping and alloy conditions establish a wide composition space for materials design, where small-gap semiconductors for optical and electronic applications would enjoy a greater freedom when tuning their material parameters such as spin-orbit coupling, electron-phonon coupling, and gap size.

\section{Methods}
All density functional theory calculations were performed using the Perdew-Burke-Ernzerhof parametrization of the generalized gradient approximation exchange-correlation functional \cite{Perdew1996, Perdew1997} and projector augmented wave potentials \cite{Joubert1999, Blochl1994}, all carried out using the Vienna Ab-initio Simulation Package \cite{Kresse1996}. Energy cutoff for plane-wave expansion was 520~eV. Force thresholds for structural relaxations were 0.02~eV/\AA. Brillouin zone sampling  were kept equivalent to a $12\times12\times1$ k-point grid for a $1\times1$ SiC surface unit cell. The DFT-D3 correction \cite{Grimme2010} was applied to all structures since graphene was included. 

\bibliographystyle{apsrev4-1}
\bibliography{Survey} 

\begin{thebibliography}{45}%
\makeatletter
\providecommand \@ifxundefined [1]{%
 \@ifx{#1\undefined}
}%
\providecommand \@ifnum [1]{%
 \ifnum #1\expandafter \@firstoftwo
 \else \expandafter \@secondoftwo
 \fi
}%
\providecommand \@ifx [1]{%
 \ifx #1\expandafter \@firstoftwo
 \else \expandafter \@secondoftwo
 \fi
}%
\providecommand \natexlab [1]{#1}%
\providecommand \enquote  [1]{``#1''}%
\providecommand \bibnamefont  [1]{#1}%
\providecommand \bibfnamefont [1]{#1}%
\providecommand \citenamefont [1]{#1}%
\providecommand \href@noop [0]{\@secondoftwo}%
\providecommand \href [0]{\begingroup \@sanitize@url \@href}%
\providecommand \@href[1]{\@@startlink{#1}\@@href}%
\providecommand \@@href[1]{\endgroup#1\@@endlink}%
\providecommand \@sanitize@url [0]{\catcode `\\12\catcode `\$12\catcode
  `\&12\catcode `\#12\catcode `\^12\catcode `\_12\catcode `\%12\relax}%
\providecommand \@@startlink[1]{}%
\providecommand \@@endlink[0]{}%
\providecommand \url  [0]{\begingroup\@sanitize@url \@url }%
\providecommand \@url [1]{\endgroup\@href {#1}{\urlprefix }}%
\providecommand \urlprefix  [0]{URL }%
\providecommand \Eprint [0]{\href }%
\providecommand \doibase [0]{http://dx.doi.org/}%
\providecommand \selectlanguage [0]{\@gobble}%
\providecommand \bibinfo  [0]{\@secondoftwo}%
\providecommand \bibfield  [0]{\@secondoftwo}%
\providecommand \translation [1]{[#1]}%
\providecommand \BibitemOpen [0]{}%
\providecommand \bibitemStop [0]{}%
\providecommand \bibitemNoStop [0]{.\EOS\space}%
\providecommand \EOS [0]{\spacefactor3000\relax}%
\providecommand \BibitemShut  [1]{\csname bibitem#1\endcsname}%
\let\auto@bib@innerbib\@empty
\bibitem [{\citenamefont {Kochat}\ \emph {et~al.}(2018)\citenamefont {Kochat},
  \citenamefont {Samanta}, \citenamefont {Zhang}, \citenamefont {Bhowmick},
  \citenamefont {Manimunda}, \citenamefont {Asif}, \citenamefont {Stender},
  \citenamefont {Vajtai}, \citenamefont {Singh}, \citenamefont {Tiwary},\ and\
  \citenamefont {Ajayan}}]{Kochat2018}%
  \BibitemOpen
  \bibfield  {author} {\bibinfo {author} {\bibfnamefont {V.}~\bibnamefont
  {Kochat}}, \bibinfo {author} {\bibfnamefont {A.}~\bibnamefont {Samanta}},
  \bibinfo {author} {\bibfnamefont {Y.}~\bibnamefont {Zhang}}, \bibinfo
  {author} {\bibfnamefont {S.}~\bibnamefont {Bhowmick}}, \bibinfo {author}
  {\bibfnamefont {P.}~\bibnamefont {Manimunda}}, \bibinfo {author}
  {\bibfnamefont {S.~A.~S.}\ \bibnamefont {Asif}}, \bibinfo {author}
  {\bibfnamefont {A.~S.}\ \bibnamefont {Stender}}, \bibinfo {author}
  {\bibfnamefont {R.}~\bibnamefont {Vajtai}}, \bibinfo {author} {\bibfnamefont
  {A.~K.}\ \bibnamefont {Singh}}, \bibinfo {author} {\bibfnamefont {C.~S.}\
  \bibnamefont {Tiwary}}, \ and\ \bibinfo {author} {\bibfnamefont {P.~M.}\
  \bibnamefont {Ajayan}},\ }\href {\doibase 10.1126/sciadv.1701373} {\bibfield
  {journal} {\bibinfo  {journal} {Sci. Adv.}\ }\textbf {\bibinfo {volume}
  {4}},\ \bibinfo {pages} {e1701373} (\bibinfo {year} {2018})}\BibitemShut
  {NoStop}%
\bibitem [{\citenamefont {Xing}\ \emph {et~al.}(2015)\citenamefont {Xing},
  \citenamefont {Zhang}, \citenamefont {Fu}, \citenamefont {Liu}, \citenamefont
  {Sun}, \citenamefont {Peng}, \citenamefont {Wang}, \citenamefont {Lin},
  \citenamefont {Ma}, \citenamefont {Xue}, \citenamefont {Wang},\ and\
  \citenamefont {Xie}}]{Xing2015}%
  \BibitemOpen
  \bibfield  {author} {\bibinfo {author} {\bibfnamefont {Y.}~\bibnamefont
  {Xing}}, \bibinfo {author} {\bibfnamefont {H.~M.}\ \bibnamefont {Zhang}},
  \bibinfo {author} {\bibfnamefont {H.~L.}\ \bibnamefont {Fu}}, \bibinfo
  {author} {\bibfnamefont {H.}~\bibnamefont {Liu}}, \bibinfo {author}
  {\bibfnamefont {Y.}~\bibnamefont {Sun}}, \bibinfo {author} {\bibfnamefont
  {J.~P.}\ \bibnamefont {Peng}}, \bibinfo {author} {\bibfnamefont
  {F.}~\bibnamefont {Wang}}, \bibinfo {author} {\bibfnamefont {X.}~\bibnamefont
  {Lin}}, \bibinfo {author} {\bibfnamefont {X.~C.}\ \bibnamefont {Ma}},
  \bibinfo {author} {\bibfnamefont {Q.~K.}\ \bibnamefont {Xue}}, \bibinfo
  {author} {\bibfnamefont {J.}~\bibnamefont {Wang}}, \ and\ \bibinfo {author}
  {\bibfnamefont {X.~C.}\ \bibnamefont {Xie}},\ }\href {\doibase
  10.1126/science.aaa7154} {\bibfield  {journal} {\bibinfo  {journal}
  {Science}\ }\textbf {\bibinfo {volume} {350}},\ \bibinfo {pages} {542}
  (\bibinfo {year} {2015})}\BibitemShut {NoStop}%
\bibitem [{\citenamefont {Yan}\ \emph {et~al.}(2019)\citenamefont {Yan},
  \citenamefont {Khalsa}, \citenamefont {Schaefer}, \citenamefont {Jarjour},
  \citenamefont {Rouvimov}, \citenamefont {Nowack}, \citenamefont {Xing},\ and\
  \citenamefont {Jena}}]{Yan2019}%
  \BibitemOpen
  \bibfield  {author} {\bibinfo {author} {\bibfnamefont {R.}~\bibnamefont
  {Yan}}, \bibinfo {author} {\bibfnamefont {G.}~\bibnamefont {Khalsa}},
  \bibinfo {author} {\bibfnamefont {B.~T.}\ \bibnamefont {Schaefer}}, \bibinfo
  {author} {\bibfnamefont {A.}~\bibnamefont {Jarjour}}, \bibinfo {author}
  {\bibfnamefont {S.}~\bibnamefont {Rouvimov}}, \bibinfo {author}
  {\bibfnamefont {K.~C.}\ \bibnamefont {Nowack}}, \bibinfo {author}
  {\bibfnamefont {H.~G.}\ \bibnamefont {Xing}}, \ and\ \bibinfo {author}
  {\bibfnamefont {D.}~\bibnamefont {Jena}},\ }\href {\doibase
  10.7567/1882-0786/aaff89} {\bibfield  {journal} {\bibinfo  {journal} {Appl.
  Phys. Express}\ }\textbf {\bibinfo {volume} {12}},\ \bibinfo {pages} {023008}
  (\bibinfo {year} {2019})}\BibitemShut {NoStop}%
\bibitem [{\citenamefont {Xi}\ \emph {et~al.}(2016)\citenamefont {Xi},
  \citenamefont {Wang}, \citenamefont {Zhao}, \citenamefont {Park},
  \citenamefont {Law}, \citenamefont {Berger}, \citenamefont {Forr{\'{o}}},
  \citenamefont {Shan},\ and\ \citenamefont {Mak}}]{Xi2016}%
  \BibitemOpen
  \bibfield  {author} {\bibinfo {author} {\bibfnamefont {X.}~\bibnamefont
  {Xi}}, \bibinfo {author} {\bibfnamefont {Z.}~\bibnamefont {Wang}}, \bibinfo
  {author} {\bibfnamefont {W.}~\bibnamefont {Zhao}}, \bibinfo {author}
  {\bibfnamefont {J.~H.}\ \bibnamefont {Park}}, \bibinfo {author}
  {\bibfnamefont {K.~T.}\ \bibnamefont {Law}}, \bibinfo {author} {\bibfnamefont
  {H.}~\bibnamefont {Berger}}, \bibinfo {author} {\bibfnamefont
  {L.}~\bibnamefont {Forr{\'{o}}}}, \bibinfo {author} {\bibfnamefont
  {J.}~\bibnamefont {Shan}}, \ and\ \bibinfo {author} {\bibfnamefont {K.~F.}\
  \bibnamefont {Mak}},\ }\href {\doibase 10.1038/nphys3538} {\bibfield
  {journal} {\bibinfo  {journal} {Nat. Phys.}\ }\textbf {\bibinfo {volume}
  {12}},\ \bibinfo {pages} {139} (\bibinfo {year} {2016})}\BibitemShut
  {NoStop}%
\bibitem [{\citenamefont {da~Jornada}\ \emph {et~al.}(2020)\citenamefont
  {da~Jornada}, \citenamefont {Xian}, \citenamefont {Rubio},\ and\
  \citenamefont {Louie}}]{DaJornada2020}%
  \BibitemOpen
  \bibfield  {author} {\bibinfo {author} {\bibfnamefont {F.~H.}\ \bibnamefont
  {da~Jornada}}, \bibinfo {author} {\bibfnamefont {L.}~\bibnamefont {Xian}},
  \bibinfo {author} {\bibfnamefont {A.}~\bibnamefont {Rubio}}, \ and\ \bibinfo
  {author} {\bibfnamefont {S.~G.}\ \bibnamefont {Louie}},\ }\href {\doibase
  10.1038/s41467-020-14826-8} {\bibfield  {journal} {\bibinfo  {journal} {Nat.
  Commun.}\ }\textbf {\bibinfo {volume} {11}},\ \bibinfo {pages} {1013}
  (\bibinfo {year} {2020})}\BibitemShut {NoStop}%
\bibitem [{\citenamefont {Briggs}\ \emph {et~al.}(2020)\citenamefont {Briggs},
  \citenamefont {Bersch}, \citenamefont {Wang}, \citenamefont {Jiang},
  \citenamefont {Koch}, \citenamefont {Nayir}, \citenamefont {Wang},
  \citenamefont {Kolmer}, \citenamefont {Ko}, \citenamefont {{De La Fuente
  Duran}}, \citenamefont {Subramanian}, \citenamefont {Dong}, \citenamefont
  {Shallenberger}, \citenamefont {Fu}, \citenamefont {Zou}, \citenamefont
  {Chuang}, \citenamefont {Gai}, \citenamefont {Li}, \citenamefont {Bostwick},
  \citenamefont {Jozwiak}, \citenamefont {Chang}, \citenamefont {Rotenberg},
  \citenamefont {Zhu}, \citenamefont {van Duin}, \citenamefont {Crespi},\ and\
  \citenamefont {Robinson}}]{Briggs2020}%
  \BibitemOpen
  \bibfield  {author} {\bibinfo {author} {\bibfnamefont {N.}~\bibnamefont
  {Briggs}}, \bibinfo {author} {\bibfnamefont {B.}~\bibnamefont {Bersch}},
  \bibinfo {author} {\bibfnamefont {Y.}~\bibnamefont {Wang}}, \bibinfo {author}
  {\bibfnamefont {J.}~\bibnamefont {Jiang}}, \bibinfo {author} {\bibfnamefont
  {R.~J.}\ \bibnamefont {Koch}}, \bibinfo {author} {\bibfnamefont
  {N.}~\bibnamefont {Nayir}}, \bibinfo {author} {\bibfnamefont
  {K.}~\bibnamefont {Wang}}, \bibinfo {author} {\bibfnamefont {M.}~\bibnamefont
  {Kolmer}}, \bibinfo {author} {\bibfnamefont {W.}~\bibnamefont {Ko}}, \bibinfo
  {author} {\bibfnamefont {A.}~\bibnamefont {{De La Fuente Duran}}}, \bibinfo
  {author} {\bibfnamefont {S.}~\bibnamefont {Subramanian}}, \bibinfo {author}
  {\bibfnamefont {C.}~\bibnamefont {Dong}}, \bibinfo {author} {\bibfnamefont
  {J.}~\bibnamefont {Shallenberger}}, \bibinfo {author} {\bibfnamefont
  {M.}~\bibnamefont {Fu}}, \bibinfo {author} {\bibfnamefont {Q.}~\bibnamefont
  {Zou}}, \bibinfo {author} {\bibfnamefont {Y.~W.}\ \bibnamefont {Chuang}},
  \bibinfo {author} {\bibfnamefont {Z.}~\bibnamefont {Gai}}, \bibinfo {author}
  {\bibfnamefont {A.~P.}\ \bibnamefont {Li}}, \bibinfo {author} {\bibfnamefont
  {A.}~\bibnamefont {Bostwick}}, \bibinfo {author} {\bibfnamefont
  {C.}~\bibnamefont {Jozwiak}}, \bibinfo {author} {\bibfnamefont {C.~Z.}\
  \bibnamefont {Chang}}, \bibinfo {author} {\bibfnamefont {E.}~\bibnamefont
  {Rotenberg}}, \bibinfo {author} {\bibfnamefont {J.}~\bibnamefont {Zhu}},
  \bibinfo {author} {\bibfnamefont {A.~C.}\ \bibnamefont {van Duin}}, \bibinfo
  {author} {\bibfnamefont {V.}~\bibnamefont {Crespi}}, \ and\ \bibinfo {author}
  {\bibfnamefont {J.~A.}\ \bibnamefont {Robinson}},\ }\href {\doibase
  10.1038/s41563-020-0631-x} {\bibfield  {journal} {\bibinfo  {journal} {Nat.
  Mater.}\ }\textbf {\bibinfo {volume} {19}},\ \bibinfo {pages} {637} (\bibinfo
  {year} {2020})}\BibitemShut {NoStop}%
\bibitem [{\citenamefont {Rosenzweig}\ and\ \citenamefont
  {Starke}(2020)}]{Rosenzweig2020}%
  \BibitemOpen
  \bibfield  {author} {\bibinfo {author} {\bibfnamefont {P.}~\bibnamefont
  {Rosenzweig}}\ and\ \bibinfo {author} {\bibfnamefont {U.}~\bibnamefont
  {Starke}},\ }\href {\doibase 10.1103/PhysRevB.101.201407} {\bibfield
  {journal} {\bibinfo  {journal} {Phys. Rev. B}\ }\textbf {\bibinfo {volume}
  {101}},\ \bibinfo {pages} {201407} (\bibinfo {year} {2020})}\BibitemShut
  {NoStop}%
\bibitem [{\citenamefont {Forti}\ \emph {et~al.}(2016)\citenamefont {Forti},
  \citenamefont {St{\"{o}}hr}, \citenamefont {Zakharov}, \citenamefont
  {Coletti}, \citenamefont {Emtsev},\ and\ \citenamefont {Starke}}]{Forti2016}%
  \BibitemOpen
  \bibfield  {author} {\bibinfo {author} {\bibfnamefont {S.}~\bibnamefont
  {Forti}}, \bibinfo {author} {\bibfnamefont {A.}~\bibnamefont {St{\"{o}}hr}},
  \bibinfo {author} {\bibfnamefont {A.~A.}\ \bibnamefont {Zakharov}}, \bibinfo
  {author} {\bibfnamefont {C.}~\bibnamefont {Coletti}}, \bibinfo {author}
  {\bibfnamefont {K.~V.}\ \bibnamefont {Emtsev}}, \ and\ \bibinfo {author}
  {\bibfnamefont {U.}~\bibnamefont {Starke}},\ }\href {\doibase
  10.1088/2053-1583/3/3/035003} {\bibfield  {journal} {\bibinfo  {journal} {2D
  Mater.}\ }\textbf {\bibinfo {volume} {3}},\ \bibinfo {pages} {035003}
  (\bibinfo {year} {2016})}\BibitemShut {NoStop}%
\bibitem [{\citenamefont {Kotsakidis}\ \emph {et~al.}(2020)\citenamefont
  {Kotsakidis}, \citenamefont {Grubi{\v{s}}i{\'{c}}-{\v{C}}abo}, \citenamefont
  {Yin}, \citenamefont {Tadich}, \citenamefont {Myers-Ward}, \citenamefont
  {Dejarld}, \citenamefont {Pavunny}, \citenamefont {Currie}, \citenamefont
  {Daniels}, \citenamefont {Liu}, \citenamefont {Edmonds}, \citenamefont
  {Medhekar}, \citenamefont {Gaskill}, \citenamefont {{V{\'{a}}zquez De
  Parga}},\ and\ \citenamefont {Fuhrer}}]{Kotsakidis2020}%
  \BibitemOpen
  \bibfield  {author} {\bibinfo {author} {\bibfnamefont {J.~C.}\ \bibnamefont
  {Kotsakidis}}, \bibinfo {author} {\bibfnamefont {A.}~\bibnamefont
  {Grubi{\v{s}}i{\'{c}}-{\v{C}}abo}}, \bibinfo {author} {\bibfnamefont
  {Y.}~\bibnamefont {Yin}}, \bibinfo {author} {\bibfnamefont {A.}~\bibnamefont
  {Tadich}}, \bibinfo {author} {\bibfnamefont {R.~L.}\ \bibnamefont
  {Myers-Ward}}, \bibinfo {author} {\bibfnamefont {M.}~\bibnamefont {Dejarld}},
  \bibinfo {author} {\bibfnamefont {S.~P.}\ \bibnamefont {Pavunny}}, \bibinfo
  {author} {\bibfnamefont {M.}~\bibnamefont {Currie}}, \bibinfo {author}
  {\bibfnamefont {K.~M.}\ \bibnamefont {Daniels}}, \bibinfo {author}
  {\bibfnamefont {C.}~\bibnamefont {Liu}}, \bibinfo {author} {\bibfnamefont
  {M.~T.}\ \bibnamefont {Edmonds}}, \bibinfo {author} {\bibfnamefont {N.~V.}\
  \bibnamefont {Medhekar}}, \bibinfo {author} {\bibfnamefont {D.~K.}\
  \bibnamefont {Gaskill}}, \bibinfo {author} {\bibfnamefont {A.~L.}\
  \bibnamefont {{V{\'{a}}zquez De Parga}}}, \ and\ \bibinfo {author}
  {\bibfnamefont {M.~S.}\ \bibnamefont {Fuhrer}},\ }\href {\doibase
  10.1021/acs.chemmater.0c01729} {\bibfield  {journal} {\bibinfo  {journal}
  {Chem. Mater.}\ }\textbf {\bibinfo {volume} {32}},\ \bibinfo {pages} {6464}
  (\bibinfo {year} {2020})},\ \Eprint {http://arxiv.org/abs/2004.01383}
  {arXiv:2004.01383} \BibitemShut {NoStop}%
\bibitem [{\citenamefont {Grubi{\v{s}}i{\'{c}}-{\v{C}}abo}\ \emph
  {et~al.}(2020)\citenamefont {Grubi{\v{s}}i{\'{c}}-{\v{C}}abo}, \citenamefont
  {Kotsakidis}, \citenamefont {Yin}, \citenamefont {Tadich}, \citenamefont
  {Haldon}, \citenamefont {Solari}, \citenamefont {di~Bernardo}, \citenamefont
  {Daniels}, \citenamefont {Riley}, \citenamefont {Huwald}, \citenamefont
  {Edmonds}, \citenamefont {Myers-Ward}, \citenamefont {Medhekar},
  \citenamefont {Gaskill},\ and\ \citenamefont {Fuhrer}}]{Grubisic-Cabo2020}%
  \BibitemOpen
  \bibfield  {author} {\bibinfo {author} {\bibfnamefont {A.}~\bibnamefont
  {Grubi{\v{s}}i{\'{c}}-{\v{C}}abo}}, \bibinfo {author} {\bibfnamefont {J.~C.}\
  \bibnamefont {Kotsakidis}}, \bibinfo {author} {\bibfnamefont
  {Y.}~\bibnamefont {Yin}}, \bibinfo {author} {\bibfnamefont {A.}~\bibnamefont
  {Tadich}}, \bibinfo {author} {\bibfnamefont {M.}~\bibnamefont {Haldon}},
  \bibinfo {author} {\bibfnamefont {S.}~\bibnamefont {Solari}}, \bibinfo
  {author} {\bibfnamefont {I.}~\bibnamefont {di~Bernardo}}, \bibinfo {author}
  {\bibfnamefont {K.~M.}\ \bibnamefont {Daniels}}, \bibinfo {author}
  {\bibfnamefont {J.}~\bibnamefont {Riley}}, \bibinfo {author} {\bibfnamefont
  {E.}~\bibnamefont {Huwald}}, \bibinfo {author} {\bibfnamefont {M.~T.}\
  \bibnamefont {Edmonds}}, \bibinfo {author} {\bibfnamefont {R.}~\bibnamefont
  {Myers-Ward}}, \bibinfo {author} {\bibfnamefont {N.~V.}\ \bibnamefont
  {Medhekar}}, \bibinfo {author} {\bibfnamefont {D.~K.}\ \bibnamefont
  {Gaskill}}, \ and\ \bibinfo {author} {\bibfnamefont {M.~S.}\ \bibnamefont
  {Fuhrer}},\ }\href {http://arxiv.org/abs/2005.02670} {\  (\bibinfo {year}
  {2020})},\ \Eprint {http://arxiv.org/abs/2005.02670} {arXiv:2005.02670}
  \BibitemShut {NoStop}%
\bibitem [{\citenamefont {Forti}\ \emph {et~al.}(2020)\citenamefont {Forti},
  \citenamefont {Link}, \citenamefont {St{\"{o}}hr}, \citenamefont {Niu},
  \citenamefont {Zakharov}, \citenamefont {Coletti},\ and\ \citenamefont
  {Starke}}]{Forti2020}%
  \BibitemOpen
  \bibfield  {author} {\bibinfo {author} {\bibfnamefont {S.}~\bibnamefont
  {Forti}}, \bibinfo {author} {\bibfnamefont {S.}~\bibnamefont {Link}},
  \bibinfo {author} {\bibfnamefont {A.}~\bibnamefont {St{\"{o}}hr}}, \bibinfo
  {author} {\bibfnamefont {Y.}~\bibnamefont {Niu}}, \bibinfo {author}
  {\bibfnamefont {A.~A.}\ \bibnamefont {Zakharov}}, \bibinfo {author}
  {\bibfnamefont {C.}~\bibnamefont {Coletti}}, \ and\ \bibinfo {author}
  {\bibfnamefont {U.}~\bibnamefont {Starke}},\ }\href {\doibase
  10.1038/s41467-020-15683-1} {\bibfield  {journal} {\bibinfo  {journal} {Nat.
  Commun.}\ }\textbf {\bibinfo {volume} {11}},\ \bibinfo {pages} {2236}
  (\bibinfo {year} {2020})}\BibitemShut {NoStop}%
\bibitem [{\citenamefont {Xia}\ \emph {et~al.}(2014)\citenamefont {Xia},
  \citenamefont {Johansson}, \citenamefont {Niu}, \citenamefont {Zakharov},
  \citenamefont {Janz{\'{e}}n}, \citenamefont {Virojanadara},\ and\
  \citenamefont {Janze}}]{Xia2014}%
  \BibitemOpen
  \bibfield  {author} {\bibinfo {author} {\bibfnamefont {C.}~\bibnamefont
  {Xia}}, \bibinfo {author} {\bibfnamefont {L.~I.}\ \bibnamefont {Johansson}},
  \bibinfo {author} {\bibfnamefont {Y.}~\bibnamefont {Niu}}, \bibinfo {author}
  {\bibfnamefont {A.~A.}\ \bibnamefont {Zakharov}}, \bibinfo {author}
  {\bibfnamefont {E.}~\bibnamefont {Janz{\'{e}}n}}, \bibinfo {author}
  {\bibfnamefont {C.}~\bibnamefont {Virojanadara}}, \ and\ \bibinfo {author}
  {\bibfnamefont {E.}~\bibnamefont {Janze}},\ }\href {\doibase
  10.1016/j.carbon.2014.08.027} {\bibfield  {journal} {\bibinfo  {journal}
  {Carbon}\ }\textbf {\bibinfo {volume} {79}},\ \bibinfo {pages} {631}
  (\bibinfo {year} {2014})}\BibitemShut {NoStop}%
\bibitem [{\citenamefont {Yurtsever}\ \emph {et~al.}(2016)\citenamefont
  {Yurtsever}, \citenamefont {Onoda}, \citenamefont {Iimori}, \citenamefont
  {Niki}, \citenamefont {Miyamachi}, \citenamefont {Abe}, \citenamefont
  {Mizuno}, \citenamefont {Tanaka}, \citenamefont {Komori},\ and\ \citenamefont
  {Sugimoto}}]{Yurtsever2016}%
  \BibitemOpen
  \bibfield  {author} {\bibinfo {author} {\bibfnamefont {A.}~\bibnamefont
  {Yurtsever}}, \bibinfo {author} {\bibfnamefont {J.}~\bibnamefont {Onoda}},
  \bibinfo {author} {\bibfnamefont {T.}~\bibnamefont {Iimori}}, \bibinfo
  {author} {\bibfnamefont {K.}~\bibnamefont {Niki}}, \bibinfo {author}
  {\bibfnamefont {T.}~\bibnamefont {Miyamachi}}, \bibinfo {author}
  {\bibfnamefont {M.}~\bibnamefont {Abe}}, \bibinfo {author} {\bibfnamefont
  {S.}~\bibnamefont {Mizuno}}, \bibinfo {author} {\bibfnamefont
  {S.}~\bibnamefont {Tanaka}}, \bibinfo {author} {\bibfnamefont
  {F.}~\bibnamefont {Komori}}, \ and\ \bibinfo {author} {\bibfnamefont
  {Y.}~\bibnamefont {Sugimoto}},\ }\href {\doibase 10.1002/smll.201600666}
  {\bibfield  {journal} {\bibinfo  {journal} {Small}\ }\textbf {\bibinfo
  {volume} {12}},\ \bibinfo {pages} {3956} (\bibinfo {year}
  {2016})}\BibitemShut {NoStop}%
\bibitem [{\citenamefont {Gao}\ \emph {et~al.}(2012)\citenamefont {Gao},
  \citenamefont {Gao}, \citenamefont {Chang}, \citenamefont {Chen},
  \citenamefont {Liu}, \citenamefont {Xie}, \citenamefont {He}, \citenamefont
  {Ma}, \citenamefont {Zhang},\ and\ \citenamefont {Liu}}]{Gao2012}%
  \BibitemOpen
  \bibfield  {author} {\bibinfo {author} {\bibfnamefont {T.}~\bibnamefont
  {Gao}}, \bibinfo {author} {\bibfnamefont {Y.}~\bibnamefont {Gao}}, \bibinfo
  {author} {\bibfnamefont {C.}~\bibnamefont {Chang}}, \bibinfo {author}
  {\bibfnamefont {Y.}~\bibnamefont {Chen}}, \bibinfo {author} {\bibfnamefont
  {M.}~\bibnamefont {Liu}}, \bibinfo {author} {\bibfnamefont {S.}~\bibnamefont
  {Xie}}, \bibinfo {author} {\bibfnamefont {K.}~\bibnamefont {He}}, \bibinfo
  {author} {\bibfnamefont {X.}~\bibnamefont {Ma}}, \bibinfo {author}
  {\bibfnamefont {Y.}~\bibnamefont {Zhang}}, \ and\ \bibinfo {author}
  {\bibfnamefont {Z.}~\bibnamefont {Liu}},\ }\href {\doibase 10.1021/nn302303n}
  {\bibfield  {journal} {\bibinfo  {journal} {ACS Nano}\ }\textbf {\bibinfo
  {volume} {6}},\ \bibinfo {pages} {6562} (\bibinfo {year} {2012})}\BibitemShut
  {NoStop}%
\bibitem [{\citenamefont {Reis}\ \emph {et~al.}(2017)\citenamefont {Reis},
  \citenamefont {Li}, \citenamefont {Dudy}, \citenamefont {Bauernfeind},
  \citenamefont {Glass}, \citenamefont {Hanke}, \citenamefont {Thomale},
  \citenamefont {Sch{\"{a}}fer},\ and\ \citenamefont {Claessen}}]{Reis2017}%
  \BibitemOpen
  \bibfield  {author} {\bibinfo {author} {\bibfnamefont {F.}~\bibnamefont
  {Reis}}, \bibinfo {author} {\bibfnamefont {G.}~\bibnamefont {Li}}, \bibinfo
  {author} {\bibfnamefont {L.}~\bibnamefont {Dudy}}, \bibinfo {author}
  {\bibfnamefont {M.}~\bibnamefont {Bauernfeind}}, \bibinfo {author}
  {\bibfnamefont {S.}~\bibnamefont {Glass}}, \bibinfo {author} {\bibfnamefont
  {W.}~\bibnamefont {Hanke}}, \bibinfo {author} {\bibfnamefont
  {R.}~\bibnamefont {Thomale}}, \bibinfo {author} {\bibfnamefont
  {J.}~\bibnamefont {Sch{\"{a}}fer}}, \ and\ \bibinfo {author} {\bibfnamefont
  {R.}~\bibnamefont {Claessen}},\ }\href {\doibase 10.1126/science.aai8142}
  {\bibfield  {journal} {\bibinfo  {journal} {Science}\ }\textbf {\bibinfo
  {volume} {357}},\ \bibinfo {pages} {287} (\bibinfo {year}
  {2017})}\BibitemShut {NoStop}%
\bibitem [{\citenamefont {Yaji}\ \emph {et~al.}(2019)\citenamefont {Yaji},
  \citenamefont {Visikovskiy}, \citenamefont {Iimori}, \citenamefont {Kuroda},
  \citenamefont {Hayashi}, \citenamefont {Kajiwara}, \citenamefont {Tanaka},
  \citenamefont {Komori},\ and\ \citenamefont {Shin}}]{Yaji2019}%
  \BibitemOpen
  \bibfield  {author} {\bibinfo {author} {\bibfnamefont {K.}~\bibnamefont
  {Yaji}}, \bibinfo {author} {\bibfnamefont {A.}~\bibnamefont {Visikovskiy}},
  \bibinfo {author} {\bibfnamefont {T.}~\bibnamefont {Iimori}}, \bibinfo
  {author} {\bibfnamefont {K.}~\bibnamefont {Kuroda}}, \bibinfo {author}
  {\bibfnamefont {S.}~\bibnamefont {Hayashi}}, \bibinfo {author} {\bibfnamefont
  {T.}~\bibnamefont {Kajiwara}}, \bibinfo {author} {\bibfnamefont
  {S.}~\bibnamefont {Tanaka}}, \bibinfo {author} {\bibfnamefont
  {F.}~\bibnamefont {Komori}}, \ and\ \bibinfo {author} {\bibfnamefont
  {S.}~\bibnamefont {Shin}},\ }\href {\doibase 10.1103/PhysRevLett.122.126403}
  {\bibfield  {journal} {\bibinfo  {journal} {Phys. Rev. Lett.}\ }\textbf
  {\bibinfo {volume} {122}},\ \bibinfo {pages} {126403} (\bibinfo {year}
  {2019})}\BibitemShut {NoStop}%
\bibitem [{\citenamefont {Li}\ \emph {et~al.}(2013)\citenamefont {Li},
  \citenamefont {Tang}, \citenamefont {Chen}, \citenamefont {Wu}, \citenamefont
  {Gu}, \citenamefont {Fang}, \citenamefont {Zhang},\ and\ \citenamefont
  {Duan}}]{Li2013a}%
  \BibitemOpen
  \bibfield  {author} {\bibinfo {author} {\bibfnamefont {Y.}~\bibnamefont
  {Li}}, \bibinfo {author} {\bibfnamefont {P.}~\bibnamefont {Tang}}, \bibinfo
  {author} {\bibfnamefont {P.}~\bibnamefont {Chen}}, \bibinfo {author}
  {\bibfnamefont {J.}~\bibnamefont {Wu}}, \bibinfo {author} {\bibfnamefont
  {B.~L.}\ \bibnamefont {Gu}}, \bibinfo {author} {\bibfnamefont
  {Y.}~\bibnamefont {Fang}}, \bibinfo {author} {\bibfnamefont {S.~B.}\
  \bibnamefont {Zhang}}, \ and\ \bibinfo {author} {\bibfnamefont
  {W.}~\bibnamefont {Duan}},\ }\href {\doibase 10.1103/PhysRevB.87.245127}
  {\bibfield  {journal} {\bibinfo  {journal} {Phys. Rev. B}\ }\textbf {\bibinfo
  {volume} {87}},\ \bibinfo {pages} {245127} (\bibinfo {year}
  {2013})}\BibitemShut {NoStop}%
\bibitem [{\citenamefont {McChesney}\ \emph {et~al.}(2010)\citenamefont
  {McChesney}, \citenamefont {Bostwick}, \citenamefont {Ohta}, \citenamefont
  {Seyller}, \citenamefont {Horn}, \citenamefont {Gonz{\'{a}}lez},\ and\
  \citenamefont {Rotenberg}}]{McChesney2010}%
  \BibitemOpen
  \bibfield  {author} {\bibinfo {author} {\bibfnamefont {J.~L.}\ \bibnamefont
  {McChesney}}, \bibinfo {author} {\bibfnamefont {A.}~\bibnamefont {Bostwick}},
  \bibinfo {author} {\bibfnamefont {T.}~\bibnamefont {Ohta}}, \bibinfo {author}
  {\bibfnamefont {T.}~\bibnamefont {Seyller}}, \bibinfo {author} {\bibfnamefont
  {K.}~\bibnamefont {Horn}}, \bibinfo {author} {\bibfnamefont {J.}~\bibnamefont
  {Gonz{\'{a}}lez}}, \ and\ \bibinfo {author} {\bibfnamefont {E.}~\bibnamefont
  {Rotenberg}},\ }\href {\doibase 10.1103/PhysRevLett.104.136803} {\bibfield
  {journal} {\bibinfo  {journal} {Phys. Rev. Lett.}\ }\textbf {\bibinfo
  {volume} {104}},\ \bibinfo {pages} {136803} (\bibinfo {year}
  {2010})}\BibitemShut {NoStop}%
\bibitem [{\citenamefont {Link}\ \emph {et~al.}(2019)\citenamefont {Link},
  \citenamefont {Forti}, \citenamefont {St{\"{o}}hr}, \citenamefont
  {K{\"{u}}ster}, \citenamefont {R{\"{o}}sner}, \citenamefont {Hirschmeier},
  \citenamefont {Chen}, \citenamefont {Avila}, \citenamefont {Asensio},
  \citenamefont {Zakharov}, \citenamefont {Wehling}, \citenamefont
  {Lichtenstein}, \citenamefont {Katsnelson},\ and\ \citenamefont
  {Starke}}]{Link2019}%
  \BibitemOpen
  \bibfield  {author} {\bibinfo {author} {\bibfnamefont {S.}~\bibnamefont
  {Link}}, \bibinfo {author} {\bibfnamefont {S.}~\bibnamefont {Forti}},
  \bibinfo {author} {\bibfnamefont {A.}~\bibnamefont {St{\"{o}}hr}}, \bibinfo
  {author} {\bibfnamefont {K.}~\bibnamefont {K{\"{u}}ster}}, \bibinfo {author}
  {\bibfnamefont {M.}~\bibnamefont {R{\"{o}}sner}}, \bibinfo {author}
  {\bibfnamefont {D.}~\bibnamefont {Hirschmeier}}, \bibinfo {author}
  {\bibfnamefont {C.}~\bibnamefont {Chen}}, \bibinfo {author} {\bibfnamefont
  {J.}~\bibnamefont {Avila}}, \bibinfo {author} {\bibfnamefont {M.~C.}\
  \bibnamefont {Asensio}}, \bibinfo {author} {\bibfnamefont {A.~A.}\
  \bibnamefont {Zakharov}}, \bibinfo {author} {\bibfnamefont {T.~O.}\
  \bibnamefont {Wehling}}, \bibinfo {author} {\bibfnamefont {A.~I.}\
  \bibnamefont {Lichtenstein}}, \bibinfo {author} {\bibfnamefont {M.~I.}\
  \bibnamefont {Katsnelson}}, \ and\ \bibinfo {author} {\bibfnamefont
  {U.}~\bibnamefont {Starke}},\ }\href {\doibase 10.1103/PhysRevB.100.121407}
  {\bibfield  {journal} {\bibinfo  {journal} {Phys. Rev. B}\ }\textbf {\bibinfo
  {volume} {100}},\ \bibinfo {pages} {121407} (\bibinfo {year}
  {2019})}\BibitemShut {NoStop}%
\bibitem [{\citenamefont {Xu}\ \emph {et~al.}(2017)\citenamefont {Xu},
  \citenamefont {Zhang}, \citenamefont {Shen}, \citenamefont {Cheng},
  \citenamefont {Schwingenschl{\"{o}}gl},\ and\ \citenamefont
  {Huang}}]{Xu2017}%
  \BibitemOpen
  \bibfield  {author} {\bibinfo {author} {\bibfnamefont {Z.}~\bibnamefont
  {Xu}}, \bibinfo {author} {\bibfnamefont {Q.}~\bibnamefont {Zhang}}, \bibinfo
  {author} {\bibfnamefont {Q.}~\bibnamefont {Shen}}, \bibinfo {author}
  {\bibfnamefont {Y.}~\bibnamefont {Cheng}}, \bibinfo {author} {\bibfnamefont
  {U.}~\bibnamefont {Schwingenschl{\"{o}}gl}}, \ and\ \bibinfo {author}
  {\bibfnamefont {W.}~\bibnamefont {Huang}},\ }\href {\doibase
  10.1039/C7TC03799F} {\bibfield  {journal} {\bibinfo  {journal} {J. Mater.
  Chem. C}\ }\textbf {\bibinfo {volume} {5}},\ \bibinfo {pages} {10427}
  (\bibinfo {year} {2017})}\BibitemShut {NoStop}%
\bibitem [{\citenamefont {Zhang}\ \emph {et~al.}(2010)\citenamefont {Zhang},
  \citenamefont {Cheng}, \citenamefont {Li}, \citenamefont {Sun}, \citenamefont
  {Wang}, \citenamefont {Zhu}, \citenamefont {He}, \citenamefont {Wang},
  \citenamefont {Ma}, \citenamefont {Chen}, \citenamefont {Wang}, \citenamefont
  {Liu}, \citenamefont {Lin}, \citenamefont {Jia},\ and\ \citenamefont
  {Xue}}]{Zhang2010}%
  \BibitemOpen
  \bibfield  {author} {\bibinfo {author} {\bibfnamefont {T.}~\bibnamefont
  {Zhang}}, \bibinfo {author} {\bibfnamefont {P.}~\bibnamefont {Cheng}},
  \bibinfo {author} {\bibfnamefont {W.~J.}\ \bibnamefont {Li}}, \bibinfo
  {author} {\bibfnamefont {Y.~J.}\ \bibnamefont {Sun}}, \bibinfo {author}
  {\bibfnamefont {G.}~\bibnamefont {Wang}}, \bibinfo {author} {\bibfnamefont
  {X.~G.}\ \bibnamefont {Zhu}}, \bibinfo {author} {\bibfnamefont
  {K.}~\bibnamefont {He}}, \bibinfo {author} {\bibfnamefont {L.}~\bibnamefont
  {Wang}}, \bibinfo {author} {\bibfnamefont {X.}~\bibnamefont {Ma}}, \bibinfo
  {author} {\bibfnamefont {X.}~\bibnamefont {Chen}}, \bibinfo {author}
  {\bibfnamefont {Y.}~\bibnamefont {Wang}}, \bibinfo {author} {\bibfnamefont
  {Y.}~\bibnamefont {Liu}}, \bibinfo {author} {\bibfnamefont {H.~Q.}\
  \bibnamefont {Lin}}, \bibinfo {author} {\bibfnamefont {J.~F.}\ \bibnamefont
  {Jia}}, \ and\ \bibinfo {author} {\bibfnamefont {Q.~K.}\ \bibnamefont
  {Xue}},\ }\href {\doibase 10.1038/nphys1499} {\bibfield  {journal} {\bibinfo
  {journal} {Nat. Phys.}\ }\textbf {\bibinfo {volume} {6}},\ \bibinfo {pages}
  {104} (\bibinfo {year} {2010})}\BibitemShut {NoStop}%
\bibitem [{\citenamefont {Zhang}\ \emph {et~al.}(2015)\citenamefont {Zhang},
  \citenamefont {Sun}, \citenamefont {Li}, \citenamefont {Peng}, \citenamefont
  {Song}, \citenamefont {Xing}, \citenamefont {Zhang}, \citenamefont {Guan},
  \citenamefont {Li}, \citenamefont {Zhao}, \citenamefont {Ji}, \citenamefont
  {Wang}, \citenamefont {He}, \citenamefont {Chen}, \citenamefont {Gu},
  \citenamefont {Ling}, \citenamefont {Tian}, \citenamefont {Li}, \citenamefont
  {Xie}, \citenamefont {Liu}, \citenamefont {Yang}, \citenamefont {Xue},
  \citenamefont {Wang},\ and\ \citenamefont {Ma}}]{Zhang2015a}%
  \BibitemOpen
  \bibfield  {author} {\bibinfo {author} {\bibfnamefont {H.~M.}\ \bibnamefont
  {Zhang}}, \bibinfo {author} {\bibfnamefont {Y.}~\bibnamefont {Sun}}, \bibinfo
  {author} {\bibfnamefont {W.}~\bibnamefont {Li}}, \bibinfo {author}
  {\bibfnamefont {J.~P.}\ \bibnamefont {Peng}}, \bibinfo {author}
  {\bibfnamefont {C.~L.}\ \bibnamefont {Song}}, \bibinfo {author}
  {\bibfnamefont {Y.}~\bibnamefont {Xing}}, \bibinfo {author} {\bibfnamefont
  {Q.}~\bibnamefont {Zhang}}, \bibinfo {author} {\bibfnamefont
  {J.}~\bibnamefont {Guan}}, \bibinfo {author} {\bibfnamefont {Z.}~\bibnamefont
  {Li}}, \bibinfo {author} {\bibfnamefont {Y.}~\bibnamefont {Zhao}}, \bibinfo
  {author} {\bibfnamefont {S.}~\bibnamefont {Ji}}, \bibinfo {author}
  {\bibfnamefont {L.}~\bibnamefont {Wang}}, \bibinfo {author} {\bibfnamefont
  {K.}~\bibnamefont {He}}, \bibinfo {author} {\bibfnamefont {X.}~\bibnamefont
  {Chen}}, \bibinfo {author} {\bibfnamefont {L.}~\bibnamefont {Gu}}, \bibinfo
  {author} {\bibfnamefont {L.}~\bibnamefont {Ling}}, \bibinfo {author}
  {\bibfnamefont {M.}~\bibnamefont {Tian}}, \bibinfo {author} {\bibfnamefont
  {L.}~\bibnamefont {Li}}, \bibinfo {author} {\bibfnamefont {X.~C.}\
  \bibnamefont {Xie}}, \bibinfo {author} {\bibfnamefont {J.}~\bibnamefont
  {Liu}}, \bibinfo {author} {\bibfnamefont {H.}~\bibnamefont {Yang}}, \bibinfo
  {author} {\bibfnamefont {Q.~K.}\ \bibnamefont {Xue}}, \bibinfo {author}
  {\bibfnamefont {J.}~\bibnamefont {Wang}}, \ and\ \bibinfo {author}
  {\bibfnamefont {X.}~\bibnamefont {Ma}},\ }\href {\doibase
  10.1103/PhysRevLett.114.107003} {\bibfield  {journal} {\bibinfo  {journal}
  {Phys. Rev. Lett.}\ }\textbf {\bibinfo {volume} {114}},\ \bibinfo {pages}
  {107003} (\bibinfo {year} {2015})}\BibitemShut {NoStop}%
\bibitem [{\citenamefont {Steves}\ \emph {et~al.}(2020)\citenamefont {Steves},
  \citenamefont {Wang}, \citenamefont {Briggs}, \citenamefont {Zhao},
  \citenamefont {El-Sherif}, \citenamefont {Bersch}, \citenamefont
  {Subramanian}, \citenamefont {Dong}, \citenamefont {Bowen}, \citenamefont
  {{Fuente Duran}}, \citenamefont {Nisi}, \citenamefont {Lassauni{\`{e}}re},
  \citenamefont {Wurstbauer}, \citenamefont {Bassim}, \citenamefont {Fonseca},
  \citenamefont {Robinson}, \citenamefont {Crespi}, \citenamefont {Robinson},\
  and\ \citenamefont {{Knappenberger Jr}}}]{Steves2020}%
  \BibitemOpen
  \bibfield  {author} {\bibinfo {author} {\bibfnamefont {M.~A.}\ \bibnamefont
  {Steves}}, \bibinfo {author} {\bibfnamefont {Y.}~\bibnamefont {Wang}},
  \bibinfo {author} {\bibfnamefont {N.}~\bibnamefont {Briggs}}, \bibinfo
  {author} {\bibfnamefont {T.}~\bibnamefont {Zhao}}, \bibinfo {author}
  {\bibfnamefont {H.}~\bibnamefont {El-Sherif}}, \bibinfo {author}
  {\bibfnamefont {B.~M.}\ \bibnamefont {Bersch}}, \bibinfo {author}
  {\bibfnamefont {S.}~\bibnamefont {Subramanian}}, \bibinfo {author}
  {\bibfnamefont {C.}~\bibnamefont {Dong}}, \bibinfo {author} {\bibfnamefont
  {T.}~\bibnamefont {Bowen}}, \bibinfo {author} {\bibfnamefont {A.~D.~L.}\
  \bibnamefont {{Fuente Duran}}}, \bibinfo {author} {\bibfnamefont
  {K.}~\bibnamefont {Nisi}}, \bibinfo {author} {\bibfnamefont {M.}~\bibnamefont
  {Lassauni{\`{e}}re}}, \bibinfo {author} {\bibfnamefont {U.}~\bibnamefont
  {Wurstbauer}}, \bibinfo {author} {\bibfnamefont {N.~D.}\ \bibnamefont
  {Bassim}}, \bibinfo {author} {\bibfnamefont {J.}~\bibnamefont {Fonseca}},
  \bibinfo {author} {\bibfnamefont {J.~T.}\ \bibnamefont {Robinson}}, \bibinfo
  {author} {\bibfnamefont {V.~H.}\ \bibnamefont {Crespi}}, \bibinfo {author}
  {\bibfnamefont {J.}~\bibnamefont {Robinson}}, \ and\ \bibinfo {author}
  {\bibfnamefont {K.~L.}\ \bibnamefont {{Knappenberger Jr}}},\ }\href {\doibase
  10.1021/acs.nanolett.0c03481} {\bibfield  {journal} {\bibinfo  {journal}
  {Nano Lett.}\ ,\ \bibinfo {pages} {acs.nanolett.0c03481}} (\bibinfo {year}
  {2020})}\BibitemShut {NoStop}%
\bibitem [{\citenamefont {Webb}\ \emph {et~al.}(2015)\citenamefont {Webb},
  \citenamefont {Marsiglio},\ and\ \citenamefont {Hirsch}}]{Webb2015}%
  \BibitemOpen
  \bibfield  {author} {\bibinfo {author} {\bibfnamefont {G.}~\bibnamefont
  {Webb}}, \bibinfo {author} {\bibfnamefont {F.}~\bibnamefont {Marsiglio}}, \
  and\ \bibinfo {author} {\bibfnamefont {J.}~\bibnamefont {Hirsch}},\ }\href
  {\doibase 10.1016/j.physc.2015.02.037} {\bibfield  {journal} {\bibinfo
  {journal} {Phys. C Supercond. its Appl.}\ }\textbf {\bibinfo {volume}
  {514}},\ \bibinfo {pages} {17} (\bibinfo {year} {2015})}\BibitemShut
  {NoStop}%
\bibitem [{\citenamefont {Herzfeld}(1927)}]{Herzfeld1927}%
  \BibitemOpen
  \bibfield  {author} {\bibinfo {author} {\bibfnamefont {K.~F.}\ \bibnamefont
  {Herzfeld}},\ }\href {\doibase 10.1103/PhysRev.29.701} {\bibfield  {journal}
  {\bibinfo  {journal} {Phys. Rev.}\ }\textbf {\bibinfo {volume} {29}},\
  \bibinfo {pages} {701} (\bibinfo {year} {1927})}\BibitemShut {NoStop}%
\bibitem [{\citenamefont {Dowben}(2000)}]{Dowben2000}%
  \BibitemOpen
  \bibfield  {author} {\bibinfo {author} {\bibfnamefont {P.}~\bibnamefont
  {Dowben}},\ }\href {\doibase 10.1016/S0167-5729(00)00010-8} {\bibfield
  {journal} {\bibinfo  {journal} {Surf. Sci. Rep.}\ }\textbf {\bibinfo {volume}
  {40}},\ \bibinfo {pages} {151} (\bibinfo {year} {2000})}\BibitemShut
  {NoStop}%
\bibitem [{\citenamefont {Yakovkin}(1999)}]{Yakovkin1999}%
  \BibitemOpen
  \bibfield  {author} {\bibinfo {author} {\bibfnamefont {I.}~\bibnamefont
  {Yakovkin}},\ }\href {\doibase 10.1016/S0039-6028(99)00957-7} {\bibfield
  {journal} {\bibinfo  {journal} {Surf. Sci.}\ }\textbf {\bibinfo {volume}
  {442}},\ \bibinfo {pages} {431} (\bibinfo {year} {1999})}\BibitemShut
  {NoStop}%
\bibitem [{\citenamefont {Burdett}(1994)}]{Burdett1994}%
  \BibitemOpen
  \bibfield  {author} {\bibinfo {author} {\bibfnamefont {J.~K.}\ \bibnamefont
  {Burdett}},\ }\href {\doibase 10.1039/CS9942300299} {\bibfield  {journal}
  {\bibinfo  {journal} {Chem. Soc. Rev.}\ }\textbf {\bibinfo {volume} {23}},\
  \bibinfo {pages} {299} (\bibinfo {year} {1994})}\BibitemShut {NoStop}%
\bibitem [{\citenamefont {Nevalaita}\ and\ \citenamefont
  {Koskinen}(2018)}]{Nevalaita2018}%
  \BibitemOpen
  \bibfield  {author} {\bibinfo {author} {\bibfnamefont {J.}~\bibnamefont
  {Nevalaita}}\ and\ \bibinfo {author} {\bibfnamefont {P.}~\bibnamefont
  {Koskinen}},\ }\href {\doibase 10.1103/PhysRevB.97.035411} {\bibfield
  {journal} {\bibinfo  {journal} {Phys. Rev. B}\ }\textbf {\bibinfo {volume}
  {97}},\ \bibinfo {pages} {035411} (\bibinfo {year} {2018})}\BibitemShut
  {NoStop}%
\bibitem [{\citenamefont {Ding}\ \emph {et~al.}(1991)\citenamefont {Ding},
  \citenamefont {Chan},\ and\ \citenamefont {Ho}}]{Ding1991}%
  \BibitemOpen
  \bibfield  {author} {\bibinfo {author} {\bibfnamefont {Y.~G.}\ \bibnamefont
  {Ding}}, \bibinfo {author} {\bibfnamefont {C.~T.}\ \bibnamefont {Chan}}, \
  and\ \bibinfo {author} {\bibfnamefont {K.~M.}\ \bibnamefont {Ho}},\ }\href
  {\doibase 10.1103/PhysRevLett.67.1454} {\bibfield  {journal} {\bibinfo
  {journal} {Phys. Rev. Lett.}\ }\textbf {\bibinfo {volume} {67}},\ \bibinfo
  {pages} {1454} (\bibinfo {year} {1991})}\BibitemShut {NoStop}%
\bibitem [{\citenamefont {Burdett}(1995)}]{burdett1995chemical}%
  \BibitemOpen
  \bibfield  {author} {\bibinfo {author} {\bibfnamefont {J.~K.}\ \bibnamefont
  {Burdett}},\ }\href {https://books.google.com/books?id=s5-2QgAACAAJ} {\emph
  {\bibinfo {title} {{Chemical Bonding in Solids}}}},\ Chemical Bonding in
  Solids\ (\bibinfo  {publisher} {Oxford University Press},\ \bibinfo {year}
  {1995})\BibitemShut {NoStop}%
\bibitem [{\citenamefont {Ziman}(2011)}]{ziman2011physics}%
  \BibitemOpen
  \bibfield  {author} {\bibinfo {author} {\bibfnamefont {J.~M.}\ \bibnamefont
  {Ziman}},\ }\href@noop {} {\emph {\bibinfo {title} {{The physics of
  metals}}}}\ (\bibinfo  {publisher} {Cambridge University Press},\ \bibinfo
  {year} {2011})\BibitemShut {NoStop}%
\bibitem [{\citenamefont {Tran}\ \emph {et~al.}(2016)\citenamefont {Tran},
  \citenamefont {Xu}, \citenamefont {Radhakrishnan}, \citenamefont {Winston},
  \citenamefont {Sun}, \citenamefont {Persson},\ and\ \citenamefont
  {Ong}}]{Tran2016a}%
  \BibitemOpen
  \bibfield  {author} {\bibinfo {author} {\bibfnamefont {R.}~\bibnamefont
  {Tran}}, \bibinfo {author} {\bibfnamefont {Z.}~\bibnamefont {Xu}}, \bibinfo
  {author} {\bibfnamefont {B.}~\bibnamefont {Radhakrishnan}}, \bibinfo {author}
  {\bibfnamefont {D.}~\bibnamefont {Winston}}, \bibinfo {author} {\bibfnamefont
  {W.}~\bibnamefont {Sun}}, \bibinfo {author} {\bibfnamefont {K.~A.}\
  \bibnamefont {Persson}}, \ and\ \bibinfo {author} {\bibfnamefont {S.~P.}\
  \bibnamefont {Ong}},\ }\href {\doibase 10.1038/sdata.2016.80} {\bibfield
  {journal} {\bibinfo  {journal} {Sci. Data}\ }\textbf {\bibinfo {volume}
  {3}},\ \bibinfo {pages} {160080} (\bibinfo {year} {2016})}\BibitemShut
  {NoStop}%
\bibitem [{\citenamefont {Yaney}\ and\ \citenamefont
  {Joshi}(1990)}]{Yaney1990}%
  \BibitemOpen
  \bibfield  {author} {\bibinfo {author} {\bibfnamefont {D.~L.}\ \bibnamefont
  {Yaney}}\ and\ \bibinfo {author} {\bibfnamefont {A.}~\bibnamefont {Joshi}},\
  }\href {\doibase 10.1557/JMR.1990.2197} {\bibfield  {journal} {\bibinfo
  {journal} {J. Mater. Res.}\ }\textbf {\bibinfo {volume} {5}},\ \bibinfo
  {pages} {2197} (\bibinfo {year} {1990})}\BibitemShut {NoStop}%
\bibitem [{\citenamefont {Ong}\ \emph {et~al.}(2013)\citenamefont {Ong},
  \citenamefont {Richards}, \citenamefont {Jain}, \citenamefont {Hautier},
  \citenamefont {Kocher}, \citenamefont {Cholia}, \citenamefont {Gunter},
  \citenamefont {Chevrier}, \citenamefont {Persson},\ and\ \citenamefont
  {Ceder}}]{Ong2013}%
  \BibitemOpen
  \bibfield  {author} {\bibinfo {author} {\bibfnamefont {S.~P.}\ \bibnamefont
  {Ong}}, \bibinfo {author} {\bibfnamefont {W.~D.}\ \bibnamefont {Richards}},
  \bibinfo {author} {\bibfnamefont {A.}~\bibnamefont {Jain}}, \bibinfo {author}
  {\bibfnamefont {G.}~\bibnamefont {Hautier}}, \bibinfo {author} {\bibfnamefont
  {M.}~\bibnamefont {Kocher}}, \bibinfo {author} {\bibfnamefont
  {S.}~\bibnamefont {Cholia}}, \bibinfo {author} {\bibfnamefont
  {D.}~\bibnamefont {Gunter}}, \bibinfo {author} {\bibfnamefont {V.~L.}\
  \bibnamefont {Chevrier}}, \bibinfo {author} {\bibfnamefont {K.~A.}\
  \bibnamefont {Persson}}, \ and\ \bibinfo {author} {\bibfnamefont
  {G.}~\bibnamefont {Ceder}},\ }\href {\doibase
  10.1016/j.commatsci.2012.10.028} {\bibfield  {journal} {\bibinfo  {journal}
  {Comput. Mater. Sci.}\ }\textbf {\bibinfo {volume} {68}},\ \bibinfo {pages}
  {314} (\bibinfo {year} {2013})}\BibitemShut {NoStop}%
\bibitem [{\citenamefont {Petersen}\ and\ \citenamefont
  {Hedeg{\aa}rd}(2000)}]{Petersen2000}%
  \BibitemOpen
  \bibfield  {author} {\bibinfo {author} {\bibfnamefont {L.}~\bibnamefont
  {Petersen}}\ and\ \bibinfo {author} {\bibfnamefont {P.}~\bibnamefont
  {Hedeg{\aa}rd}},\ }\href {\doibase 10.1016/S0039-6028(00)00441-6} {\bibfield
  {journal} {\bibinfo  {journal} {Surf. Sci.}\ }\textbf {\bibinfo {volume}
  {459}},\ \bibinfo {pages} {49} (\bibinfo {year} {2000})}\BibitemShut
  {NoStop}%
\bibitem [{\citenamefont {Pettifor}(1987)}]{Pettifor1987}%
  \BibitemOpen
  \bibfield  {author} {\bibinfo {author} {\bibfnamefont {D.}~\bibnamefont
  {Pettifor}},\ }in\ \href {\doibase 10.1016/S0081-1947(08)60690-6} {\emph
  {\bibinfo {booktitle} {Solid State Phys.}}},\ \bibinfo {editor} {edited by\
  \bibinfo {editor} {\bibfnamefont {H.}~\bibnamefont {Ehrenreich}}\ and\
  \bibinfo {editor} {\bibfnamefont {D.}~\bibnamefont {Turnbull}}}\ (\bibinfo
  {publisher} {Academic Press},\ \bibinfo {year} {1987})\ pp.\ \bibinfo {pages}
  {43--92}\BibitemShut {NoStop}%
\bibitem [{\citenamefont {Paxton}\ \emph {et~al.}(1997)\citenamefont {Paxton},
  \citenamefont {Methfessel},\ and\ \citenamefont {Pettifor}}]{Paxton1997}%
  \BibitemOpen
  \bibfield  {author} {\bibinfo {author} {\bibfnamefont {A.~T.}\ \bibnamefont
  {Paxton}}, \bibinfo {author} {\bibfnamefont {M.}~\bibnamefont {Methfessel}},
  \ and\ \bibinfo {author} {\bibfnamefont {D.~G.}\ \bibnamefont {Pettifor}},\
  }\href {\doibase 10.1098/rspa.1997.0080} {\bibfield  {journal} {\bibinfo
  {journal} {Proc. R. Soc. London. Ser. A Math. Phys. Eng. Sci.}\ }\textbf
  {\bibinfo {volume} {453}},\ \bibinfo {pages} {1493} (\bibinfo {year}
  {1997})}\BibitemShut {NoStop}%
\bibitem [{\citenamefont {Mizutani}(2016)}]{Mizutani2016}%
  \BibitemOpen
  \bibfield  {author} {\bibinfo {author} {\bibfnamefont {U.}~\bibnamefont
  {Mizutani}},\ }\href {\doibase 10.1201/b10324} {\emph {\bibinfo {title}
  {{Hume-Rothery Rules for Structurally Complex Alloy Phases}}}}\ (\bibinfo
  {publisher} {CRC Press},\ \bibinfo {year} {2016})\BibitemShut {NoStop}%
\bibitem [{\citenamefont {Perdew}\ \emph {et~al.}(1996)\citenamefont {Perdew},
  \citenamefont {Burke},\ and\ \citenamefont {Ernzerhof}}]{Perdew1996}%
  \BibitemOpen
  \bibfield  {author} {\bibinfo {author} {\bibfnamefont {J.~P.}\ \bibnamefont
  {Perdew}}, \bibinfo {author} {\bibfnamefont {K.}~\bibnamefont {Burke}}, \
  and\ \bibinfo {author} {\bibfnamefont {M.}~\bibnamefont {Ernzerhof}},\ }\href
  {\doibase 10.1103/PhysRevLett.77.3865} {\bibfield  {journal} {\bibinfo
  {journal} {Phys. Rev. Lett.}\ }\textbf {\bibinfo {volume} {77}},\ \bibinfo
  {pages} {3865} (\bibinfo {year} {1996})}\BibitemShut {NoStop}%
\bibitem [{\citenamefont {Perdew}\ \emph {et~al.}(1997)\citenamefont {Perdew},
  \citenamefont {Burke},\ and\ \citenamefont {Ernzerhof}}]{Perdew1997}%
  \BibitemOpen
  \bibfield  {author} {\bibinfo {author} {\bibfnamefont {J.~P.}\ \bibnamefont
  {Perdew}}, \bibinfo {author} {\bibfnamefont {K.}~\bibnamefont {Burke}}, \
  and\ \bibinfo {author} {\bibfnamefont {M.}~\bibnamefont {Ernzerhof}},\ }\href
  {\doibase 10.1103/PhysRevLett.78.1396} {\bibfield  {journal} {\bibinfo
  {journal} {Phys. Rev. Lett.}\ }\textbf {\bibinfo {volume} {78}},\ \bibinfo
  {pages} {1396} (\bibinfo {year} {1997})}\BibitemShut {NoStop}%
\bibitem [{\citenamefont {Joubert}(1999)}]{Joubert1999}%
  \BibitemOpen
  \bibfield  {author} {\bibinfo {author} {\bibfnamefont {D.}~\bibnamefont
  {Joubert}},\ }\href {\doibase 10.1103/PhysRevB.59.1758} {\bibfield  {journal}
  {\bibinfo  {journal} {Phys. Rev. B}\ }\textbf {\bibinfo {volume} {59}},\
  \bibinfo {pages} {1758} (\bibinfo {year} {1999})}\BibitemShut {NoStop}%
\bibitem [{\citenamefont {Bl{\"{o}}chl}(1994)}]{Blochl1994}%
  \BibitemOpen
  \bibfield  {author} {\bibinfo {author} {\bibfnamefont {P.~E.}\ \bibnamefont
  {Bl{\"{o}}chl}},\ }\href {\doibase 10.1103/PhysRevB.50.17953} {\bibfield
  {journal} {\bibinfo  {journal} {Phys. Rev. B}\ }\textbf {\bibinfo {volume}
  {50}},\ \bibinfo {pages} {17953} (\bibinfo {year} {1994})}\BibitemShut
  {NoStop}%
\bibitem [{\citenamefont {Kresse}\ and\ \citenamefont
  {Furthm{\"{u}}ller}(1996)}]{Kresse1996}%
  \BibitemOpen
  \bibfield  {author} {\bibinfo {author} {\bibfnamefont {G.}~\bibnamefont
  {Kresse}}\ and\ \bibinfo {author} {\bibfnamefont {J.}~\bibnamefont
  {Furthm{\"{u}}ller}},\ }\href {\doibase 10.1103/PhysRevB.54.11169} {\bibfield
   {journal} {\bibinfo  {journal} {Phys. Rev. B}\ }\textbf {\bibinfo {volume}
  {54}},\ \bibinfo {pages} {11169} (\bibinfo {year} {1996})}\BibitemShut
  {NoStop}%
\bibitem [{\citenamefont {Grimme}\ \emph {et~al.}(2010)\citenamefont {Grimme},
  \citenamefont {Antony}, \citenamefont {Ehrlich},\ and\ \citenamefont
  {Krieg}}]{Grimme2010}%
  \BibitemOpen
  \bibfield  {author} {\bibinfo {author} {\bibfnamefont {S.}~\bibnamefont
  {Grimme}}, \bibinfo {author} {\bibfnamefont {J.}~\bibnamefont {Antony}},
  \bibinfo {author} {\bibfnamefont {S.}~\bibnamefont {Ehrlich}}, \ and\
  \bibinfo {author} {\bibfnamefont {H.}~\bibnamefont {Krieg}},\ }\href
  {\doibase 10.1063/1.3382344} {\bibfield  {journal} {\bibinfo  {journal} {J.
  Chem. Phys.}\ }\textbf {\bibinfo {volume} {132}},\ \bibinfo {pages} {154104}
  (\bibinfo {year} {2010})}\BibitemShut {NoStop}%
\end{thebibliography}%
\end{document}